\documentclass[11pt]{article}
\RequirePackage[margin=1in]{geometry}

\usepackage{natbib}
\usepackage{algorithm}
\usepackage{algpseudocode}
\usepackage{etoolbox}\AtBeginEnvironment{algorithmic}{\small}
\usepackage{amsmath}
\usepackage{amssymb}
\usepackage{amsfonts}
\usepackage{bm}
\usepackage{xcolor}
\usepackage{graphicx}
\usepackage{url}

\def\R{\mathbb{R}}

\newcommand{\diag}{\text{diag}}
\newcommand{\V}[1]{{\bm{\mathbf{\MakeLowercase{#1}}}}} % vector
\newcommand{\M}[1]{{\bm{\mathbf{\MakeUppercase{#1}}}}} % matrix

\providecommand{\keywords}[1]
{
  \small	
  \textbf{\textit{Keywords:}} #1
}

\usepackage{amsthm}

\newtheoremstyle{named}{}{}{\itshape}{}{\bfseries}{.}{.5em}{\thmnote{#3}}
\theoremstyle{named}

\title{Conformal Prediction for Astronomy Data with Measurement Error}
\author{Naomi Giertych, Jonathan P Williams, Sujit Ghosh}

\begin{document}
\maketitle

\begin{abstract}
Astronomers often deal with data where the covariates and the dependent variable are measured with heteroscedastic non-Gaussian error. For instance, while \textit{TESS} and \textit{Kepler} datasets provide a wealth of information, addressing the challenges of measurement errors and systematic biases is critical for extracting reliable scientific insights and improving machine learning models' performance. Although techniques have been developed for estimating regression parameters for these data, few techniques exist to construct prediction intervals with finite sample coverage guarantees. To address this issue, we tailor the conformal prediction approach to our application. We empirically demonstrate that this method gives finite sample control over Type I error probabilities under a variety of assumptions on the measurement errors in the observed data. Further, we demonstrate how the conformal prediction method could be used for constructing prediction intervals for unobserved exoplanet masses using established broken power-law relationships between masses and radii found in the literature.
\end{abstract}

\keywords{heteroscedastic measurement error, prediction, exoplanets and satellites: fundamental parameters}

%%%%%%%%%%%%%%%%%%%%%%%%%%%%%%%%%%%%%%%%%%%%%%%%%%

%%%%%%%%%%%%%%%%% BODY OF PAPER %%%%%%%%%%%%%%%%%%

\section{Introduction}

Astronomers often work with data in which both the independent and the dependent variables are measured with non-Gaussian, heteroscedastic error where the distribution of the errors is unknown and the variances are bounded by known instrument precision \citep{chen2016probabilistic, feigelson2021twenty, kelly2012measurement, wolfgang2016probabilistic}. Further, the standard linear regression noise, often referred to as {\em intrinsic scatter} in astronomy literature or {\em error in the equation} in measurement error model literature, may also be heteroscedastic with unknown distribution.

NASA's \textit{TESS} (\textit{Transiting Exoplanet Survey Satellite}) and \textit{Kepler} missions have revolutionized the search for exoplanets, providing extensive data on their properties, including masses and radii. These data sets are hosted on the NASA Exoplanet Archive (\url{https://exoplanetarchive.ipac.caltech.edu/}), a rich repository of exoplanet-related observations and analyses. The archives primarily rely on light curves to infer planetary properties, where the key measurements are derived from the periodic dimming of a star as a planet transits across it. Radii are derived from the transit depth, which depends on accurate stellar radius measurements. Errors in stellar characterization propagate to planetary radius estimates. Masses often involve indirect methods (e.g., RV or TTV), which are prone to noise from stellar activity, instrument limitations, or insufficient observation time. Larger planets (e.g., gas giants) are more easily detected and characterized, leading to selection bias in the available data. Small planets or those orbiting faint stars often yield incomplete or less reliable measurements. Many machine learning methods assume clean, independent, and accurately labeled datasets. However, inherent measurement errors in masses and radii are often overlooked during preprocessing. These errors can bias model training and predictions, particularly for tasks such as exoplanet classification or habitability prediction. To address these challenges, recent efforts focus on incorporating uncertainty-aware methods. In this paper, we adopt the conformal prediction framework to take into account heterogeneous measurement errors and, thus, help alleviate concerns about the statistical properties of prediction sets constructed with observations containing the measurement issues aforementioned with planetary systems.

A significant reason for using the conformal prediction framework is because the methods for prediction that have been proposed in the measurement error model literature rely on one or more assumptions that astronomers cannot justify in practice. It is demonstrated in \cite{lindley1947regression} that the prediction interval derived from simple linear regression is appropriate for the observed dependent variable provided that the distributions of the covariates and the measurement errors remain the same when used for estimation and prediction. In \cite{ganse1983prediction} a prediction method is suggested when the covariate used for prediction does not come from the same distribution as that used for estimation, but the distribution of the measurement errors remains the same as homoscedastic Gaussian. The case where the true covariates and additive measurement errors are jointly Gaussian distributed with known covariance matrices is considered in \cite{fuller1987measurement}. The joint Gaussianity assumption for point predictors is relaxed in \cite{buonaccorsi2010measurement}, but the authors did not construct a prediction interval for this scenario. Somewhat completing this endeavor, suggested in \cite{jiang2024prediction} is an algorithm for constructing prediction intervals when the measurement errors are non-Gaussian but a covariate without measurement error is available. Finally, considered in \cite{carroll2009nonparametric} is nonparametric prediction when the errors in the covariates are heteroscedastic with known distributions.

However, the conformal prediction method discussed in \cite{shafer2008tutorial} and \cite{vovk2005} significantly relaxes the assumptions needed for constructing valid prediction sets. In particular, their method only requires that any ordering of the observations has the same distribution as the original set, a property known as {\em exchangeability}. We demonstrate that under a mild assumption on the errors, data typically encountered by astronomers can be transformed to exchangeable observations and used to construct prediction intervals that are robust to unknown, but bounded-variance, heteroscedastic measurement errors and possibly heteroscedastic intrinsic scatter. To the best of our knowledge, no other paper has tailored the conformal prediction method for this scenario.

The rest of this paper is organized as follows. Section \ref{sec:cp} provides an overview of the conformal prediction method for a set of exchangeable random variables. Section \ref{sec:mem} details the measurement error model considered in this paper. Section \ref{sec:methods} demonstrates that a scaled version of the residuals is robust to misspecification of the measurement error distributions, under certain assumptions, and discusses how we estimate these scaled residuals. Section \ref{sec:sims} demonstrates that the prediction intervals constructed for the measurement error model considered in this paper achieve the nominal coverage under a variety of settings. Section \ref{sec:mr} applies our methods to the piecewise mass-radius relationship, often referred to as the \textit{broken power law}, model employed by astronomers. Section \ref{sec:concl} concludes with some final remarks.

\section{Conformal Prediction}\label{sec:cp}

Conformal prediction (CP) is a method of constructing prediction sets with exact finite sample coverage with minimal assumptions on the data. Namely, the only assumption needed for CP is that the data consists of exchangeable observations. Informally, observations or a sequence of random variables are exchangeable if the joint distribution of the random variables is unchanged when the order of the random variables changes. Formally, exchangeable random variables can be defined as follows (\cite{shafer2008tutorial}):

\vspace{12pt}
\noindent
\textbf{Exchangeable Random Variables.} \hspace{2pt} \textit{The variables $z_1, \dots, z_N$ are exchangeable if for every permutation $\tau$ of the integers $1, \dots, N$, the variables $w_1, \dots, w_N$ where $w_i = z_{\tau(i)}$, have the same joint probability distribution as $z_1, \dots, z_N$.  Note, independent and identically distributed random variables are exchangeable.}
\vspace{12pt}

Assuming the data consist of exchangeable observations, the (transductive) CP algorithm builds a $(1 - \epsilon)100\%$ \textit{CP set}, $\mathcal{C}_{\epsilon}$, by collecting proposed values for the new observation which fit with the observed values based on a non-parametric estimation of the error distribution (or a transformation thereof) \citep{vovk2005, shafer2008tutorial}. In particular, the steps to construct a CP set are as follows. First, define a dissimilarity measure, known as the \textit{non-conformity measure}, between a given observation and the remaining. Second, provisionally set the value of the new (unobserved) response. Third, remove each observation and calculate it's \textit{non-conformity score} using the non-conformity measure.  Fourth, calculate the empirical p-value of the non-conformity score corresponding to the new observation. Finally, include the proposed value in the CP set if the empirical p-value is greater than $\epsilon$. We present these steps as Algorithm \ref{alg:cpi} for ease of reference, which is adapted from \cite{shafer2008tutorial}.

\begin{algorithm}
\caption{Constructing a Conformal Prediction Set}\label{alg:cpi}
    \begin{algorithmic}
        \State \textit{Input:} Old observations $z_1, \dots, z_n$
        \State \textit{Input:} Define non-conformity measure $A$
        \State \textit{Task:} Decide whether to include $z$ in the $(1-\epsilon)100\%$ prediction set $\mathcal{C}_{\epsilon}$
        \State \textit{Algorithm:}
            \begin{enumerate}
                \item Provisionally set $z_{n+1} := z$
                \item For $i = 1, \dots, n+1$, set $\alpha_i := A(z_i, \{z_1, \dots, z_{n+1}\} \backslash z_i)$
                \item Calculate $p := \frac{| \{\alpha_i: \alpha_i \geq \alpha_{n+1}, i = 1, \dots n+1\} |}{n+1}$
                \item Include $z$ in the prediction set $\mathcal{C}_{\epsilon}$ if $p > \epsilon$ 
            \end{enumerate}
    \end{algorithmic}
\end{algorithm}

As mentioned above, the CP framework produces prediction intervals with finite sample coverage. In other words, $(1 - \epsilon)100\%$ CP sets are guaranteed to contain the true value with at least $(1 - \epsilon)100\%$ probability regardless of the sample size. This is a significant advantage over other methods of constructing prediction sets that only achieve asymptotic coverage, i.e. correct coverage at very large sample sizes. This property is a direct consequence of the exchangeability of the non-conformity scores, inherited from the observations, and is formally stated in the theorem below (Proposition 1 of \citep{kuchibhotla2020exchangeability}). The proof is omitted for brevity.

\vspace{12pt}
\noindent
\textbf{Finite Sample Coverage.} \hspace{2pt} \textit{If the variables $z_1, \dots, z_N, z_{N+1}$ are exchangeable, $z_1, \dots, z_N$ is observed, and $z_{N+1}$ is a new test point then \begin{equation*}
    1-\alpha \leq P(z_{N+1} \in \mathcal{C}_{\epsilon}) \leq  1-\alpha + \frac{1}{N + 1}.
\end{equation*}}

\section{Statistical Model}\label{sec:mem}

We focus on a measurement error model frequently encountered in the astronomy community. Specifically, we assume the observed data are generated as follows:
\begin{equation}\label{eq:mem}
	\begin{split}
        y_i &= \mu_m(x_i,\V{\beta}) + v_i, \\
        \mu_m(x_i,\V{\beta}) &= \beta_0 + \V{\beta}_1^{\intercal} \Psi_m(x_i) \\
        y_i^{\text{obs}} &= y_i + w_i \\
		x_i^{\text{obs}} &= x_i + q_i,
	\end{split}
\end{equation} 
where $y_i$ and $y_i^{obs}$ are the unknown and observed scalar response, respectively, $x_i$ and $x_i^{obs}$ are the unknown and observed scalar covariates, respectively, $\V{\beta} = (\beta_0, \V{\beta}_1^\intercal)^\intercal$ is a column vector of unknown regression parameters (where $\cdot^\intercal$ indicates a transpose), $v_i$ is the intrinsic scatter in $y_i$, $w_i$ and $q_i$ are the unknown measurement errors on $y_i^{\text{obs}}$ and $x_i^{\text{obs}}$, respectively, and $i = 1, \dots, n$. The vector-valued function $\Psi(x)_m=(\psi_1(x),\ldots,\psi_m(x))^\intercal$ represents a suitable class of basis functions (e.g., orthogonal polynomials, Bernstein polynomials, splines, etc.) to be chosen specifically for our applications. If the number of basis functions $m$ is allowed to vary with the sample size $n$, the above model provides a rich class of regression models that allows for approximation and non-parametric estimation of any smooth regression function. The first two equations describe the regression relationship and the others describe the measurement error relationships. For example, $x_i$ could be the logarithm of the true radius of an exoplanet, $y_i$ the logarithm of its true mass, and $x_i^{\text{obs}}$ and $y_i^{\text{obs}}$ their observed counterparts.

As mentioned previously, it is typically assumed that the error variances are known and sometimes that the errors have a joint multivariate Gaussian distribution. However, astronomers are faced with estimating the measurement error model and forming prediction intervals when the error variances are bounded, but unknown, and may not follow a Gaussian distribution \citep{ma2021predicting, wolfgang2016probabilistic}. Specifically astronomers make the following assumptions on the errors, 
\begin{equation}
    \begin{split}
        E[(v_i, w_i, q_i) | x_i] &= 0 \\
    V[(v_i, w_i, q_i)  | x_i ] &= \text{diag}\{\sigma_{v_i}^2, \sigma_{w_i}^2, \sigma_{q_i}^2\} \\
    l_{w_i} < \sigma_{w_i}^2 &< u_{w_i} \\
    l_{q_i} < \sigma_{q_i}^2 &< u_{q_i}
    \end{split}
\end{equation} where $E[\cdot | x_i]$ and $V[\cdot | x_i]$ are the conditional expectation and variance given $x_i$, $\text{diag}{\cdot}$ is a diagonal matrix, $\{l_{w_i}, u_{w_i}, l_{q_i}, u_{q_i}\}_{i=1}^n$ are known bounds, and $\{(v_i, w_i, q_i | x_i)\}_{i=1}^n$ are assumed to be mutually independent.

TBecause the errors are heteroscedastic, the observations from model (\ref{eq:mem}) are not exchangeable, and we must make a minor assumption on the errors. For the purposes of this paper, we specifically assume $\{(v_i, w_i, q_i | x_i)\}_{i=1}^n$ are random vectors drawn from a distribution in the class of elliptical distributions which include the multivariate Gaussian, multivariate generalized normal, and multivariate t distributions \citep{cambanis1981theory, gomez1998multivariate, johnson1987multivariate}. However, as we show in Section \ref{sec:ncs}, there is no need to specify which elliptical distribution the errors are from as long as it does not change between observations. We also assume $\{x_i\}_{i=1}^n$ are mutually independent from some distribution. Below, we provide the definition of an elliptically contoured random variable as given by \cite{boente2014characterization}.

\vspace{12pt}
\noindent
\textbf{Elliptically Contoured Random Vector.} \textit{A $d \times 1$ random vector $\V{z}$ is said to have a $d$-dimensional spherical distribution if its distribution is invariant under orthogonal transformations, i.e. if $\M{O}\V{z} \sim \V{z}$ for any $d \times d$ orthogonal matrix $\M{O}$. Further, the characteristic function of $\V{z}$ is of the form $\psi_{\V{z}} (\V{t}) = \phi (\V{t}^{\intercal}\V{t})$ for $\V{t} \in \R^d$, and we denote such a spherically distributed random variable as $\V{z} \sim \mathcal{S}_d(\phi)$. Finally, if $\V{z}$ is absolutely continuous in $\R^d$, then it has a density of the form $f(\V{z}) = g(\V{z}^{\intercal} \V{z})$ for some function $g(s) \geq 0$. \\ \indent Elliptical distributions in $\R^d$ arise from affine transformations of spherically distributed random vectors in $\R^d$. Thus, $\V{x}$ has an elliptically contoured distribution if $\V{x} = \M{B}\V{z} + \V{\mu}$ where $\M{B}$ is a $d \times d$ matrix and $\V{\mu} \in \R^d$ and is denoted $\V{x} \sim EC_d(\V{\mu}, \M{\Sigma}, \phi)$ where $\M{\Sigma} = \M{B} \M{B}^{\intercal}$. Conveniently, the characteristic function of $\V{x}$ can be written as} 
\begin{equation*}
    \psi_{\V{x}}(t) = \exp(i\V{t}^{\intercal}\V{\mu}) \phi(\V{t}^{\intercal}\M{\Sigma} \V{t}).
\end{equation*} \textit{Further, if $\V{z}$ has the density $f$ above and $\M{\Sigma}$ is non-singular, then the density of $\V{x}$ is given by}
\begin{equation*}
    f(x) = |\M{\Sigma}|^{-1/2} g\bigg[(\V{x}-\V{\mu})^\intercal \M{\Sigma}^{-1} (\V{x}-\V{\mu})\bigg].
\end{equation*} \textit{For example, $g(t) = \exp(-t/2)$ gives a Gaussian distribution.}
%\vspace{12pt}

\section{Methods}\label{sec:methods}

In this section, we first detail our construction of the non-conformity score needed for model (\ref{eq:mem}) assuming the regression parameters and variances under model (\ref{eq:mem}) are known or can be estimated. We then discuss how we fit model (\ref{eq:mem}).

\subsection{Non-Conformity Score Construction} \label{sec:ncs}

In this section, we leverage the properties of elliptical distributions and the mutual independence of the errors to construct a sequence of independent and approximately identically distributed observations. We can use these transformed observations in Algorithm \ref{alg:cpi}, along with a choice of non-conformity score, to produce valid prediction intervals for the measurement error model.

Using the linear and quadratic results from \cite{cambanis1981theory} and \cite{johnson1987multivariate}, we first note that if the regression parameters and variances are given a fixed value, then the residuals under model (\ref{eq:mem}) can be written as follows:
\begin{equation}
    \begin{split}
        \begin{bmatrix} v_i \\ w_i \\ q_i \end{bmatrix} \biggm\lvert x_i &\overset{indep.}{\sim} EC_3 (0, c_{\phi}\text{diag}\{\sigma_{v_i}^2, \sigma_{w_i}^2, \sigma_{q_i}^2\}, \phi) \\
    \begin{bmatrix} \mu_m(x_i,\V{\beta}) + v_i \\ w_i \\ x_i + q_i \end{bmatrix} \biggm\lvert x_i &\overset{indep.}{\sim} EC_3\left(\begin{bmatrix} \mu_m(x_i,\V{\beta}) \\ 0 \\ x_i \end{bmatrix}, c_{\phi} \text{diag}\{\sigma_{v_i}^2, \sigma_{w_i}^2, \sigma_{q_i}^2\}, \phi \right) \\
    \begin{bmatrix} 1 & 1 & -\V{\beta}_1' \end{bmatrix} \begin{bmatrix} y_i \\ w_i \\ \Psi(x_i^{obs}) \end{bmatrix} \biggm\lvert x_i &\overset{indep.}{\approx} EC_1\big(\beta_0, c_{\phi} \eta_{i}, \phi \big) \\
    (y_i^{obs} - \mu_m(x_i^{obs}, \V{\beta})) \bigm\lvert x_i &\overset{indep.}{\approx} EC_1(0, c_{\phi} \eta_i , \phi)
    \end{split}
\end{equation}
where $\overset{indep.}{\sim}$ is read as independently distributed as, $\overset{indep.}{\approx}$ is read as independently and approximately distributed as, $c_{\phi}$ is a scalar constant determined by $\phi$, $\eta_i = \sigma_{v_i}^2 + \sigma^2_{w_i} + \V{\beta}_1^\intercal \nabla \Psi_m \nabla \Psi_m^\intercal\V{\beta}_1\sigma_{q_i}^2$, and $\nabla \Psi_m= (\psi_1'(x_i), \dots \psi_m'(x_i))^\intercal$ where $\psi_j'(x_i)$ is the derivative of $\psi_j(x_i)$. Note, the third line is obtained using the Taylor series approximation $\Psi_m(x_i^{obs}) \approx \Psi_m(x_i) + \nabla \Psi_m(x_i)q_i$. We can then produce independently and approximately identically distributed, i.e. approximately exchangeable, observations by scaling the residuals by $\sqrt{\eta_i}$.

We then need to choose a non-conformity measure. As mentioned above, \cite{shafer2008tutorial} define a non-conformity measure as a real-valued function of how unusual an observation is from any of the others; for convenience, it is often taken to be positive. For instance, the square or the absolute value of the residual may be taken as the non-conformity measure in the standard linear regression setting or the distance to the nearest neighbor with the same label may be used for supervised learning algorithms. For convenience, we choose the non-conformity measure as:
\begin{equation}\label{eq:ncs}
    r_i^{obs} = \frac{(y_i^{obs} - \mu_m(x_i^{obs},\V{\beta}))^2}{\eta_i}.
\end{equation} 

Because the scaled residuals produce approximately exchangeable random variables regardless of the specific distribution assumed for $(v_i, w_i, q_i | x_i)^\intercal$, we consider this implementation of the CP algorithm for the measurement error model to be robust to misspecification of the error distribution within the class of elliptically symmetric distributions. Note, also, that the distribution of $x_i$ can safely be ignored.

Finally, we replace the unknown values in equation (\ref{eq:ncs}) with their estimates to obtain the non-conformity scores, and use Algorithm (\ref{alg:cpi}) to construct the CP interval. The following section details how we estimate these unknown values.

\subsection{Feasible Generalized Least Squares for the Measurement Error Model} \label{sec:fglsmem}

We now detail how we estimate the components needed for constructing the non-conformity score in equation (\ref{eq:ncs}) above. First, we set $\sigma^2_{w_i}$ and $\sigma^2_{q_i}$ as the midpoint of their known bounds, denoted $\Tilde{\cdot}$, and initialize $\sigma^2_{v_i}$ to be one. We then use a slight modification of the moment-corrected estimators provided by \cite{buonaccorsi2010measurement} to initialize $\hat{\V{\beta}}$. Specifically, the moment-corrected estimator we use is
\begin{equation} \label{eq:beta_hat_mc}
    \hat{\V{\beta}}_{mc} = (\Psi(\M{X}^{obs})^{\intercal} \Psi(\M{X}^{obs}) - \M{\Sigma}_q)^{-1} \Psi(\M{X}^{obs})^{\intercal} \V{y}^{obs},
\end{equation} 
where
\begin{align*}
    \M{\Sigma}_q = \begin{bmatrix}
        0 & \V{0}_m^{\intercal} \\
        \V{0}_m & \sum_{i=1}^n \nabla \Psi_m(x_i) \nabla \Psi_m(x_i)^{\intercal} \Tilde{\sigma}_{q_i}^2
    \end{bmatrix}.
\end{align*} 
The details of deriving this estimator are provided in Appendix \ref{sec:beta_hat_deriv}.

In order to estimate the remaining components, we must specify $\Psi_m(x)$. For the purposes of this paper, we assume that $\Psi_m(x)$ represents the set of basis functions corresponding to a regression spline of order $p$ with $k$ knots $\zeta_1, \dots, \zeta_k$; specifically, 
\begin{equation}
    \mu_m(x_i, \V{\beta}) = \sum_{j=0}^p \beta_j x_i^j + \sum_{j=1}^k \beta_{p+j}(x_i - \zeta_j)_+^p
\end{equation} where $z_+ = zI(z>0)$ and $k=0$ corresponds to a continuous, non-piece-wise relationship \citep{carroll2009nonparametric}. For our purposes, we take the knot points as known. Finally, we define `region $i$' of the fitted line as the set of observations such that $\zeta_{i-1} < x_i \leq \zeta_i$.

Following, \cite{york1966least} and \cite{whiten1979multiple} we define the loss function for $x_i$ as: 
\begin{equation}
    \ell(x_i) = \frac{(y_i^{\text{obs}} - \mu_m(x_i, \hat{\V{\beta}}))^2}{\widehat{\sigma^2_{w_i} + \sigma^2_{v_i}}} + \frac{(x_i^{\text{obs}} - x_i)^2}{\Tilde{\sigma}_{q_i}}.
\end{equation} To estimate $x_i$ (and thus $\nabla \Psi_m$), we check for local minimums in each region (being sure to check the knot points) and select the one that globally minimizes $\ell(x_i)$ at the current estimates of $\V{\beta}$, $\sigma^2_{w_i} + \sigma^2_{v_i}$, and $\sigma^2_{q_i}$. We denote this final estimate of $x_i$ as $\hat{x}_i$. Next, we set $\widehat{\sigma^2_{w_i} + \sigma^2_{v_i}}$ equal to $\max\{l_{w_i},(y_i^{\text{obs}} - \mu_m(\hat{x}_i, \hat{\V{\beta}}))^2\}$. Finally, we update our estimate of $\V{\beta}$ using the generalized least squares estimate:
\begin{align*}
    \hat{\V{\beta}} = (\Psi(\M{X}^{obs})^{\intercal} \Sigma_{\eta}^{-1} \Psi(\M{X}^{obs}))^{-1} \Psi(\M{X}^{obs})^{\intercal} \Sigma_{\eta}^{-1} \V{y}^{obs},
\end{align*} 
where $\Sigma_{\eta}^{-1} = \diag(\hat{\eta}_1^{-2}, \dots, \hat{\eta}_n^{-2})$. We iterate $\hat{x}_i$, $\widehat{\sigma^2_{w_i} + \sigma^2_{v_i}}$, and $\hat{\V{\beta}}$ until convergence based on the relative difference between the predicted responses.

\section{Simulations} \label{sec:sims}

\begin{figure}[t]
    \centering
    \includegraphics[scale = 0.4]{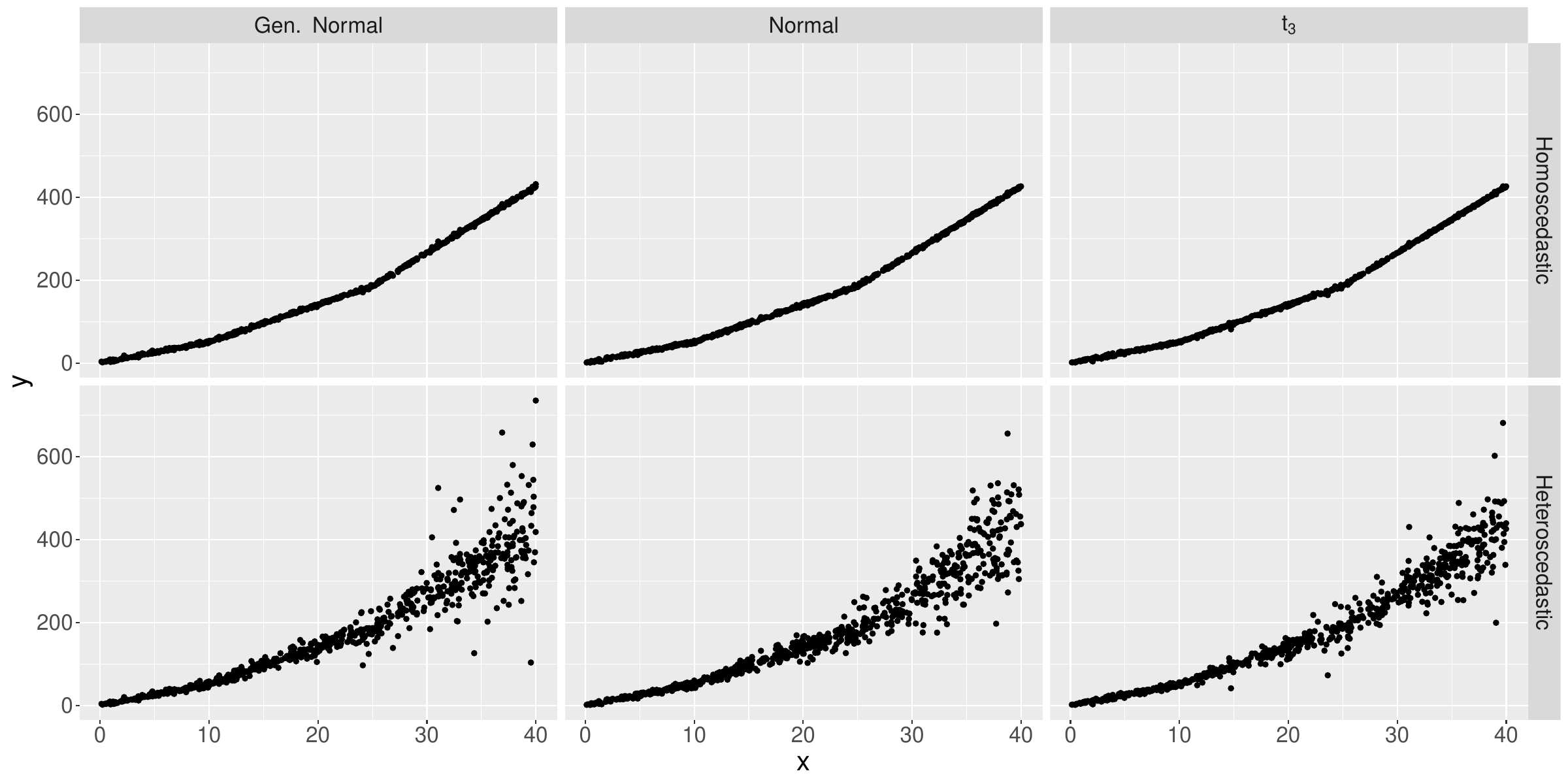}
    \caption{\small{Simulated data: Demonstration of the intrinsic scatter, based on 698 observations to make relationship clear.}}
    \label{fig:intrinsic_scatter}
\end{figure}

\begin{figure}[H]
    \centering
    \includegraphics[scale = 0.4]{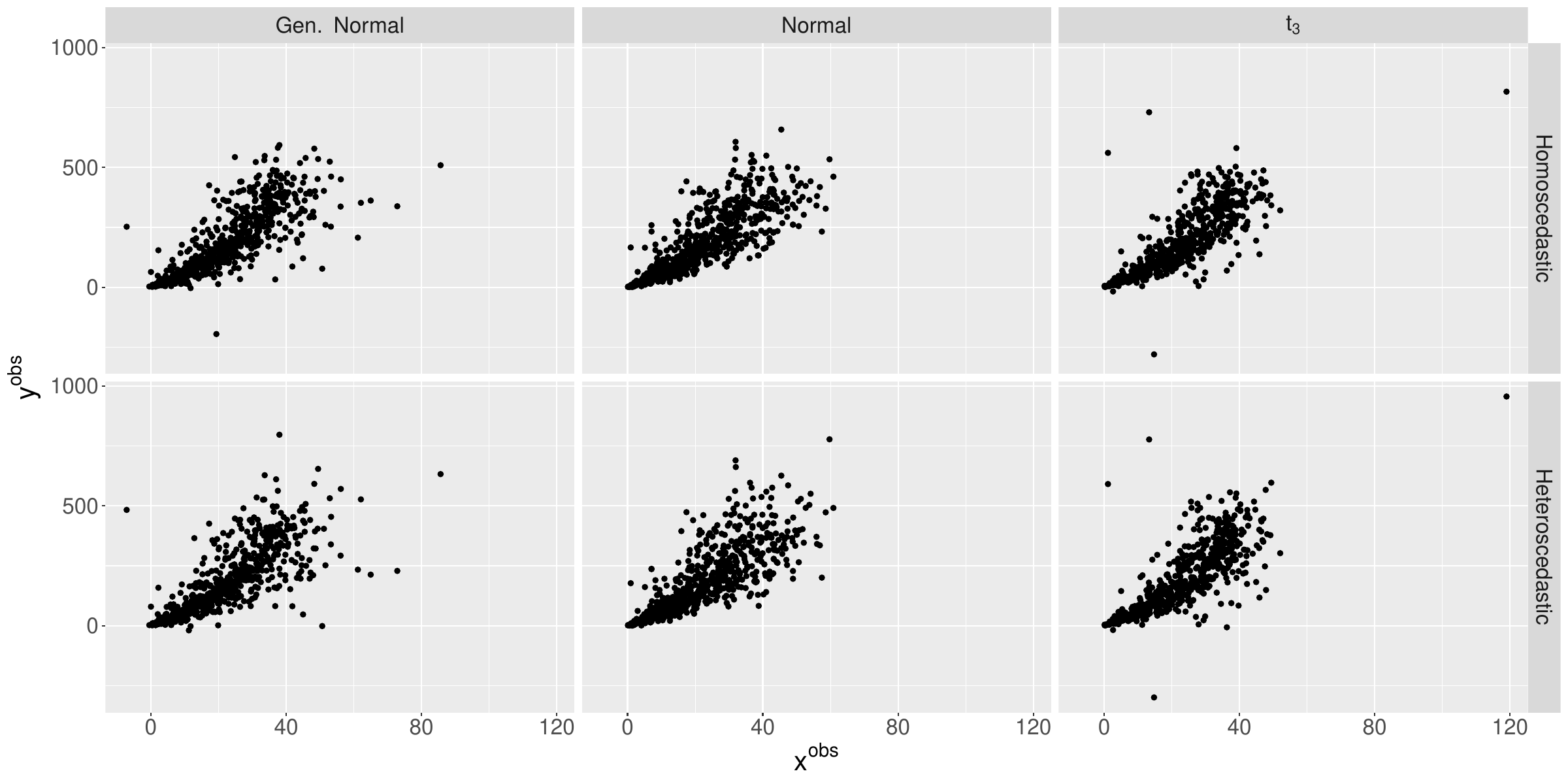}
    \caption{\small{Simulated data: Demonstration of the observed data, based on 698 observations to make relationship clear.}}
    \label{fig:obs}
\end{figure}

For our simulation study, we calculate the empirical coverages for 19 different potential prediction sets at different nominal coverages using 300 test points for each. To do this, we generate 100 training points and 19 test points (each corresponding to a test point for each prediction set) 300 times as follows. We randomly draw $x_i \sim Unif(0, 40)$ and generate $\mu_m(x_i, \V{\beta})$ from a regression spline with order one, $\V{\beta} = (1,5,4,7)$, and two knot points at 10 and 25. We then set $l_{q_i} = 0.06 x_i^2$, $u_{q_i}=0.07 x_i^2$, $l_{w_i} = 0.06 \mu_m(x_i, \V{\beta})$, $u_{w_i} = 0.07 \mu_m(x_i, \V{\beta})$. The true measurement error variances are then defined as the convex combination of the lower and upper bounds with the weight on the lower bound set to 0.3. These bounds were motivated by the observed bounds in the mass-radius data used in the real data application below. The choice of convex combination weight to determine the final measurement error variance was arbitrary but deliberately not the midpoint. We defined the true intrinsic scatter variance as $\sigma^2_{v_i} = \exp(\tau_0 + \tau_1 x_i)$ and set $\tau_0 = \log(3)$ and $\tau_1 = 0, 0.2$ to simulate homo- and hetero-scedastic intrinsic scatter. Finally, the errors are drawn from either a multivariate Gaussian distribution, a multivariate t-distribution with three degrees of freedom, or a multivariate generalized normal distribution with kurtosis parameter equal to 0.365 \citep{gomez1998multivariate}. The scale parameters of the multivariate t and generalized normal distribution were adjusted accordingly to achieve the desired variances. The t-distribution degrees of freedom parameter was chosen to demonstrate a scenario with heavy tails, and the generalized normal kurtosis parameter matches that of the generalized normal fitted to the residuals between the observed and estimated masses in \cite{ma2021predicting}, accounting for differences in notation. Figure \ref{fig:intrinsic_scatter} demonstrates the relationship between $x_i$ and $y_i$, and Figure \ref{fig:obs} demonstrates the observed relationship. It's clear that the measurement error variances obscure the true relationship greatly.

\begin{figure}[H]
    \centering
    \includegraphics[scale = 0.4]{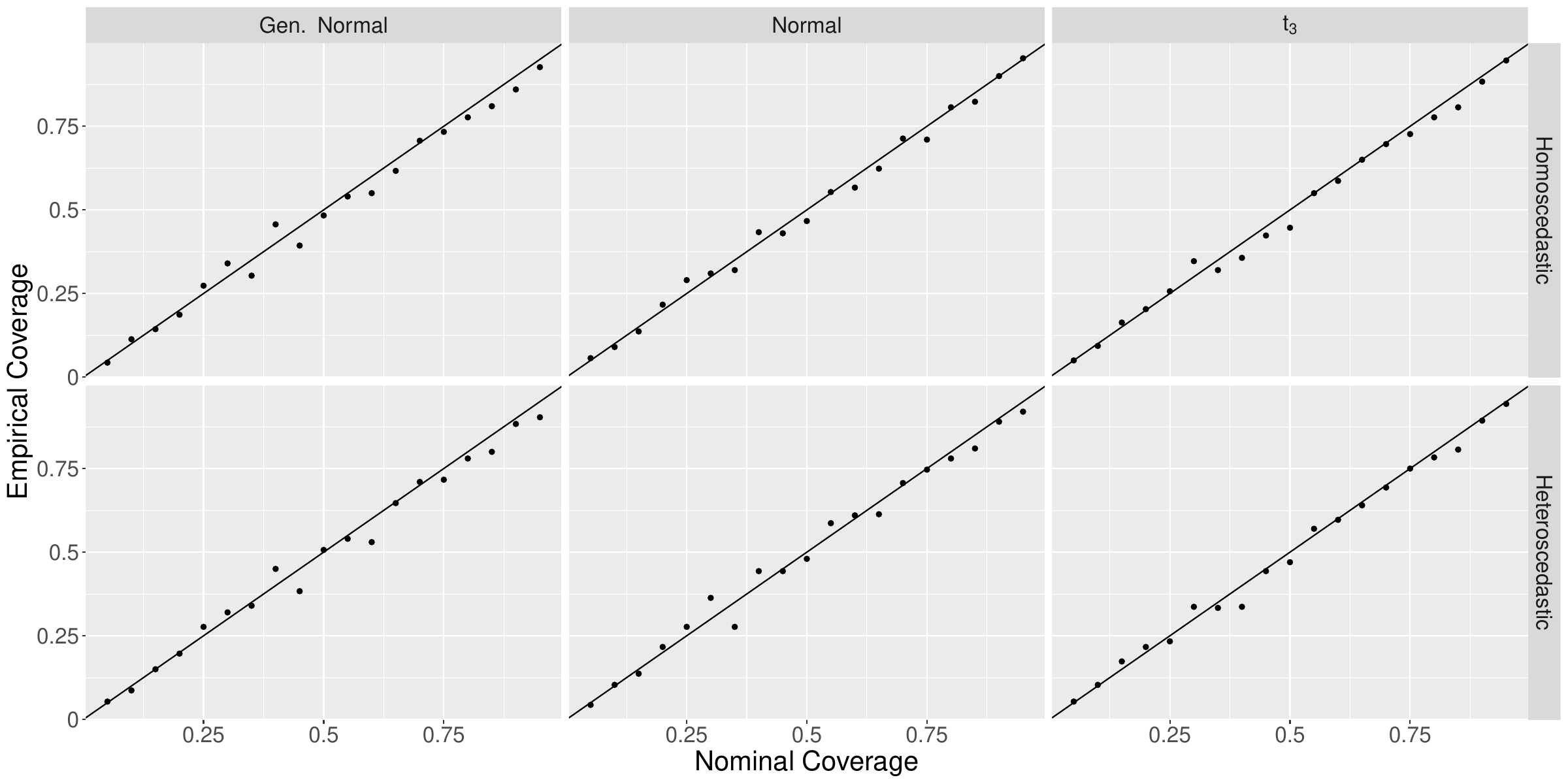}
    \caption{\small{Simulated data: Empirical versus nominal coverage across all regions.}}
    \label{fig:all_regions}
\end{figure}

\begin{figure}[H]
    \centering
    \includegraphics[scale = 0.4]{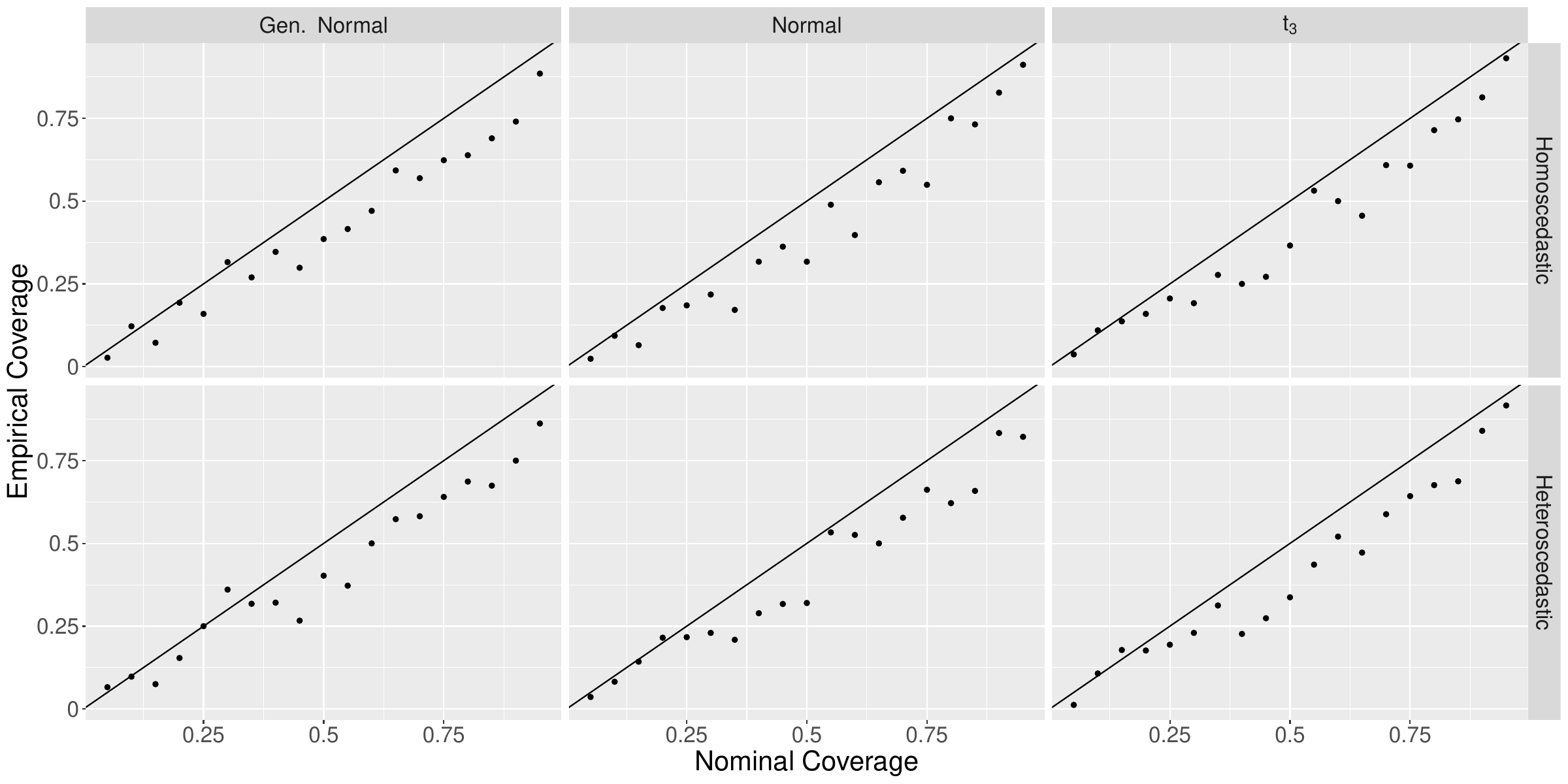}
    \caption{\small{Simulated data: Empirical versus nominal coverage for region 1 using all non-conformity scores.}}
    \label{fig:region1}
\end{figure}

For each test point, we then determine whether $y_i^{obs}$ would have been in the prediction interval using Algorithm \ref{alg:cpi}. Specifically, we set $z = y_i^{obs}$ and add the test point to the training data. Then, for each observation $i$, we remove it and calculate $\hat{\V{\beta}}_{(-i)}$ using the method described in Section \ref{sec:fglsmem} and estimate the non-conformity score $r_i^{obs}$ in equation (\ref{eq:ncs}). Note, we estimate $x_i$ and $\sigma^2_{w_i} + \sigma^2_{v_i}$ in $r_i^{obs}$ using $\hat{\V{\beta}}_{(-i)}$. Since we only care about the coverage properties of the CP intervals, we avoid calculating the entire CP set by calculating the p-value in Algorithm \ref{alg:cpi} and checking if it's larger than the specified Type I error rate $\epsilon$. As Figure \ref{fig:all_regions} shows, the empirical coverage matches the prescribed nominal coverage when examining all regions together. However, as Figure \ref{fig:region1} shows, the empirical coverages may not match the nominal coverages for each region separately.

\begin{figure}
    \centering
    \includegraphics[scale = 0.4]{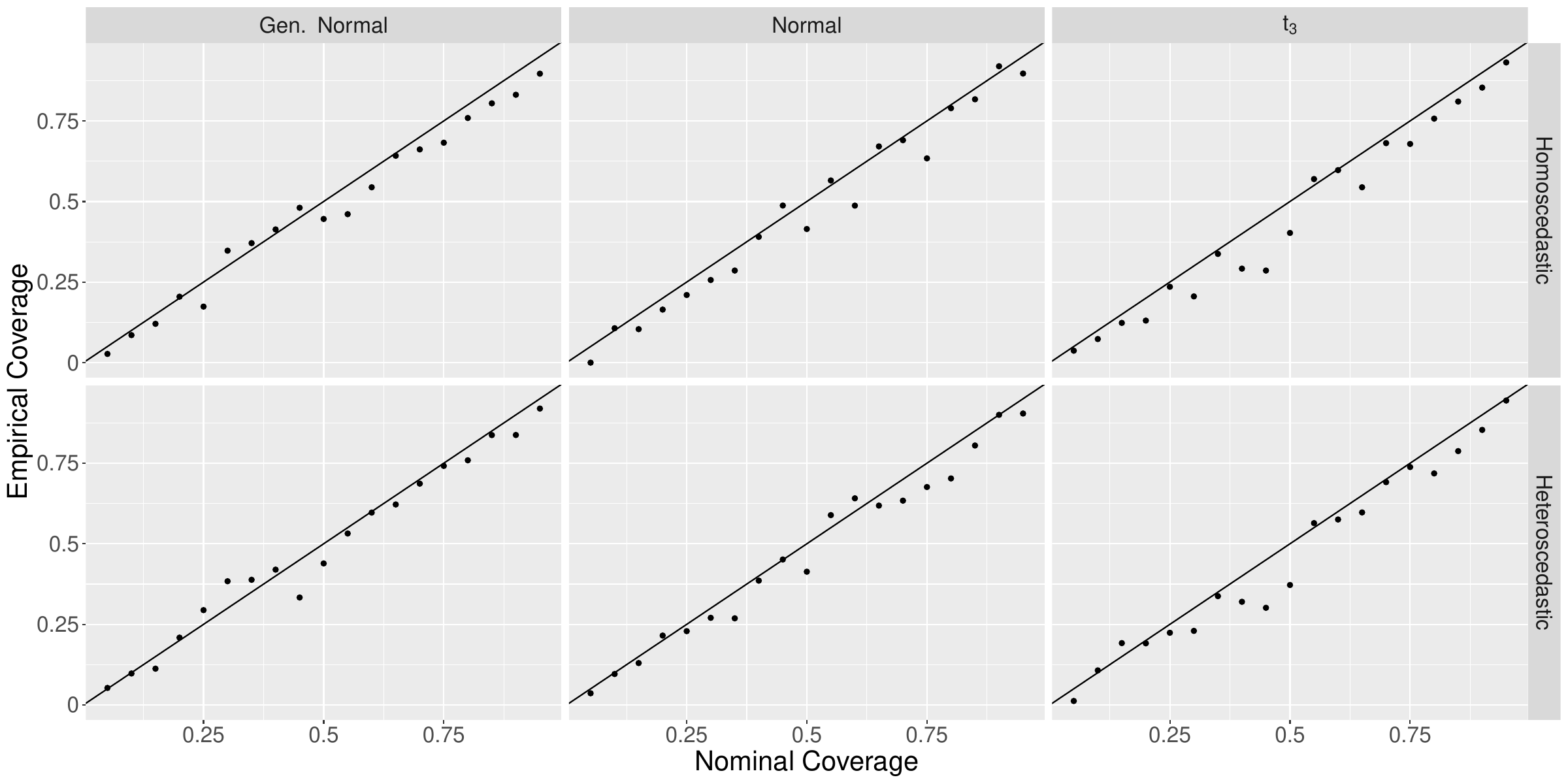}
    \caption{\small{Simulated data: Empirical versus nominal coverage for region 1 using only non-conformity scores in the same region.}}
    \label{fig:region1_adj}
\end{figure}

To correct for this, we instead calculate the p-value of the test point using only the non-conformity scores of points in the same region as that estimated for the test point. Figure \ref{fig:region1_adj} shows that the CP intervals using these adjusted p-values achieve better coverage than their non-adjusted counterparts in Figure \ref{fig:region1}. This ensures that our prediction intervals contain the true value at the expected level set by the practitioner.

\section{Real Data Application} \label{sec:mr}

While the efforts of the \textit{TESS} and \textit{Kepler} missions have significantly reduced the number of small planets without mass or radius estimates, these missions have also discovered new exoplanets that need one or both of these planetary attributes. For example, the \textit{TESS} mission has just over 3300 confirmed exoplanets with radii less than $4R_{\oplus}$ of which most, just under 2800 exoplanets, do not have mass estimates as of 11 December 2024. However, there are still significant limitations on the detectability of small planets and radial velocity follow-up is resource-intensive. Thus, an accurate prediction of the mass of a planet is crucial for targeting these radial velocity campaigns towards planets that can detected.

However, incorporating the heteroscedastic measurement errors and intrinsic scatter in a rigorous way for predictions has remained difficult. Historically, the measurement uncertainty was incorporated using a hierarchical Bayesian model formulation where the measurement errors are assumed to be normally distributed with mean zero and known variance assumed to be the midpoint of the given variance bounds \citep[e.g.,][]{wolfgang2016probabilistic, ma2019maximum, ma2021predicting}. At best, such a practice will result in unwarranted deflation of standard errors used in determining the statistical significance of parameter estimates and construction of prediction intervals. Additionally, \cite{ma2021predicting} recently demonstrated evidence that the measurement errors are likely non-Gaussian in distribution, causing the normality assumption to further misrepresent parameter estimate standard errors, statistical significance, and prediction interval widths.

In this section, we demonstrate how our methods can be used to produce valid mass predictions using the mass-radius relationship estimated in these studies. In particular, we follow \cite{ma2021predicting} and model the mass-radius (M-R) relationship as a broken power law. We chose this model because the power law relationship, $E(M) \propto CR^{\gamma}$ for some $C > 0$ and $\gamma > 0$, has been effective for characterizing the mass-radius relationship on a population level for subgroups of exoplanets (e.g. \cite{weiss2014mass} and \cite{wolfgang2016probabilistic}), and the broken power law generalizes this relationship by allowing $C$ and $\gamma$ to vary across different clusters of exoplanets (e.g. \cite{chen2016probabilistic, ma2019maximum}, and \cite{ma2021predicting}). However, we choose to break from \cite{ma2021predicting} and \cite{wolfgang2016probabilistic} by not assuming the measurement errors in the observed values to be Gaussian distributed, and instead follow \cite{chen2016probabilistic} who consider the dispersion of the measurement errors to be multiplicatively related to the true values. Specifically, our model for the mass-radius relationship is
\begin{equation}\label{eq:mrr}
    \begin{split}
        M_i &= \prod_{j=1}^J C_j R_i^{\gamma_j} \exp(v_i) I(R_i > \zeta_j) \\
        M_i^{obs} &= M_i \exp(w_i) \\
        R^{obs} &= R_i \exp(q_i)
    \end{split}
\end{equation} where $M_i$ and $R_i$ are the mass and radius of exoplanet $i$ measured in proportion to Earth's, $\zeta_j$ is the value a knot point, and $(v_i, w_i, q_i|x_i) \sim EC_3 (0, c_{\phi}\text{diag}\{\sigma_{v_i}^2, \sigma_{w_i}^2, \sigma_{q_i}^2\}, \phi)$. Since the observations differ by several orders of magnitudes, we follow \cite{ma2021predicting} and apply a base-ten logarithmic transformation. In particular, equation (\ref{eq:mrr}) can be re-written as
\begin{equation}
    \begin{split}
        \widetilde{M}_i &= \sum_{j=1}^J \bigg[ \widetilde{C_j} + \frac{\gamma_j}{\ln(10)} \widetilde{R}_i + \frac{v_i}{\ln(10)} \bigg]I(\widetilde{R_i} > \widetilde{\zeta_j}) \\
        \widetilde{M}_i^{\text{obs}} &= \widetilde{M}_i + \frac{w_i}{\ln(10)} \\
        \widetilde{R}_i^{\text{obs}} &= \widetilde{R}_i + \frac{q_i}{\ln(10)}
    \end{split}
\end{equation} where $\widetilde{\cdot}$ denotes the log base-10 of the variable. As above, $\sigma_{w_i}^2$ and $\sigma_{q_i}^2$ are unknown but bounded, and $\sigma^2_{v_i}$ is unknown. Since our main goal is prediction, we assume the knot points are known and set them to be 1.5, 3.84, 8, and 13 based on the results in \cite{ma2019maximum} and \cite{ma2021predicting}.

\begin{figure}
    \centering
    \includegraphics[scale = 0.3]{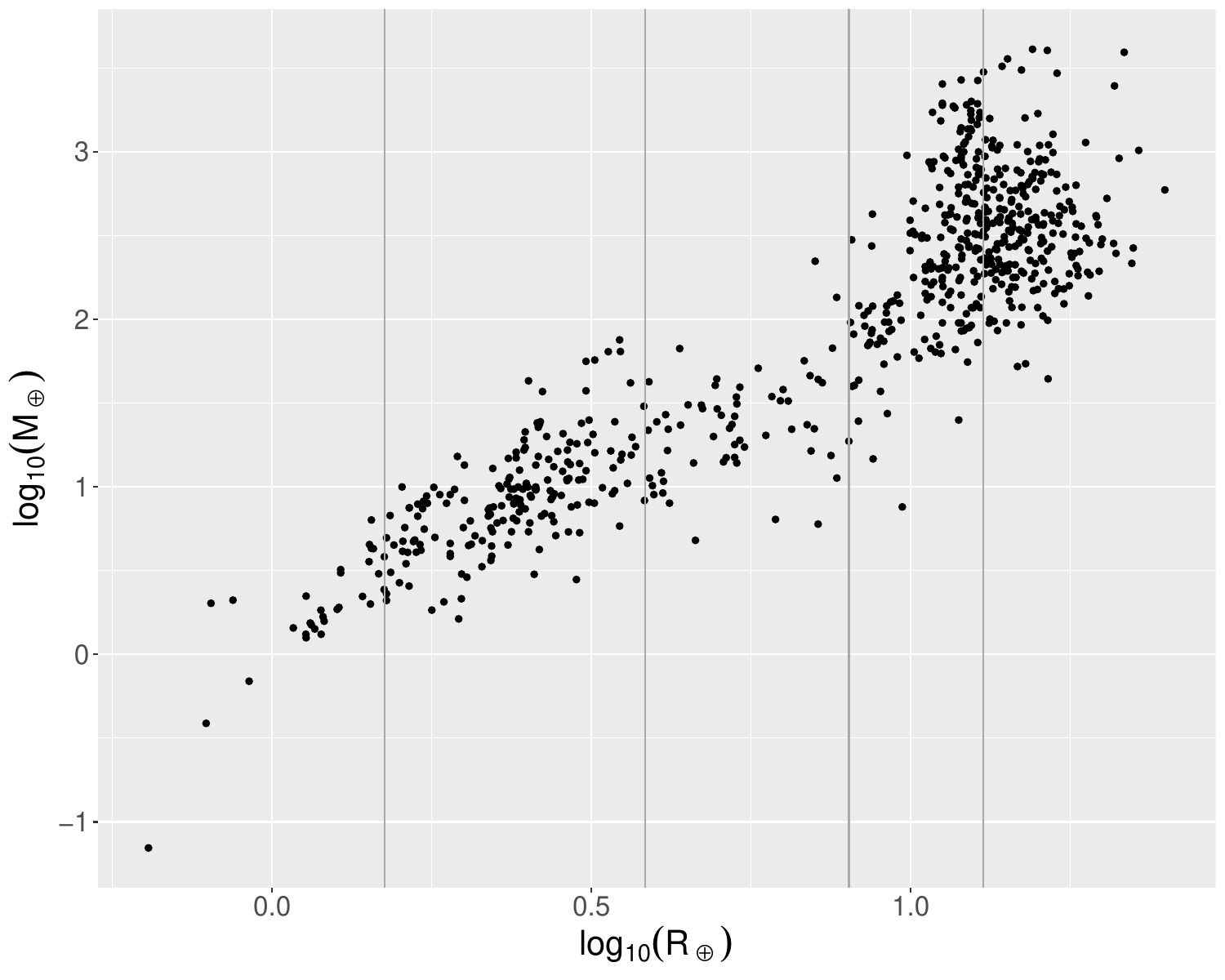}
    \caption{\small{Mass-radius relation: Random selection of 698 observations used as the training set from a total of 998 observations from the NASA Exoplanet Archive Planetary Systems Table on 4 November 2024 \citep{akeson2013nasa, ps}. Grey vertical lines indicate knot points in the fitted broken power law.}}
    \label{fig:nasa_train}
\end{figure}

The data considered here were acquired from the NASA Exoplanet Archive Planetary Systems Table on 4 November 2024 \citep{akeson2013nasa, ps}. Observed mass and radius measurements were required to come from exoplanets that were not flagged as controversial and had error variance bounds. Additionally, if an exoplanet had multiple estimates of the mass or radius, the default parameter set was used as the observed values.  Since we are only interested in the mass-radius relationship for exoplanets, we exclude brown dwarfs exhibiting deuterium fusion by introducing an upper mass boundary at $13M_J$ \citep{spiegel2011deuterium, ma2021predicting}. Finally, following \cite{wolfgang2016probabilistic} and \cite{ma2010variable}, we use the midpoint of the upper and lower variance bounds as the ``known'' value and require all measurements to have a high signal-to-noise relationship by employing a $3\sigma$ cutoff, i.e. $M_i^{\text{obs}} /\sigma_{w_i} > 3$ and $R_i^{\text{obs}} /\sigma_{q_i} > 3$. There were a total of 998 exoplanets that met those criteria; we randomly select 698 of these observations to as the training set and the remaining 300 observations are as new test points. Figure \ref{fig:nasa_train} plots the observed masses and radii in the training data along with vertical lines indicating the known knot points used in fitting the broken power law relationship.

\begin{figure}
    \centering
    \includegraphics[scale = 0.3]{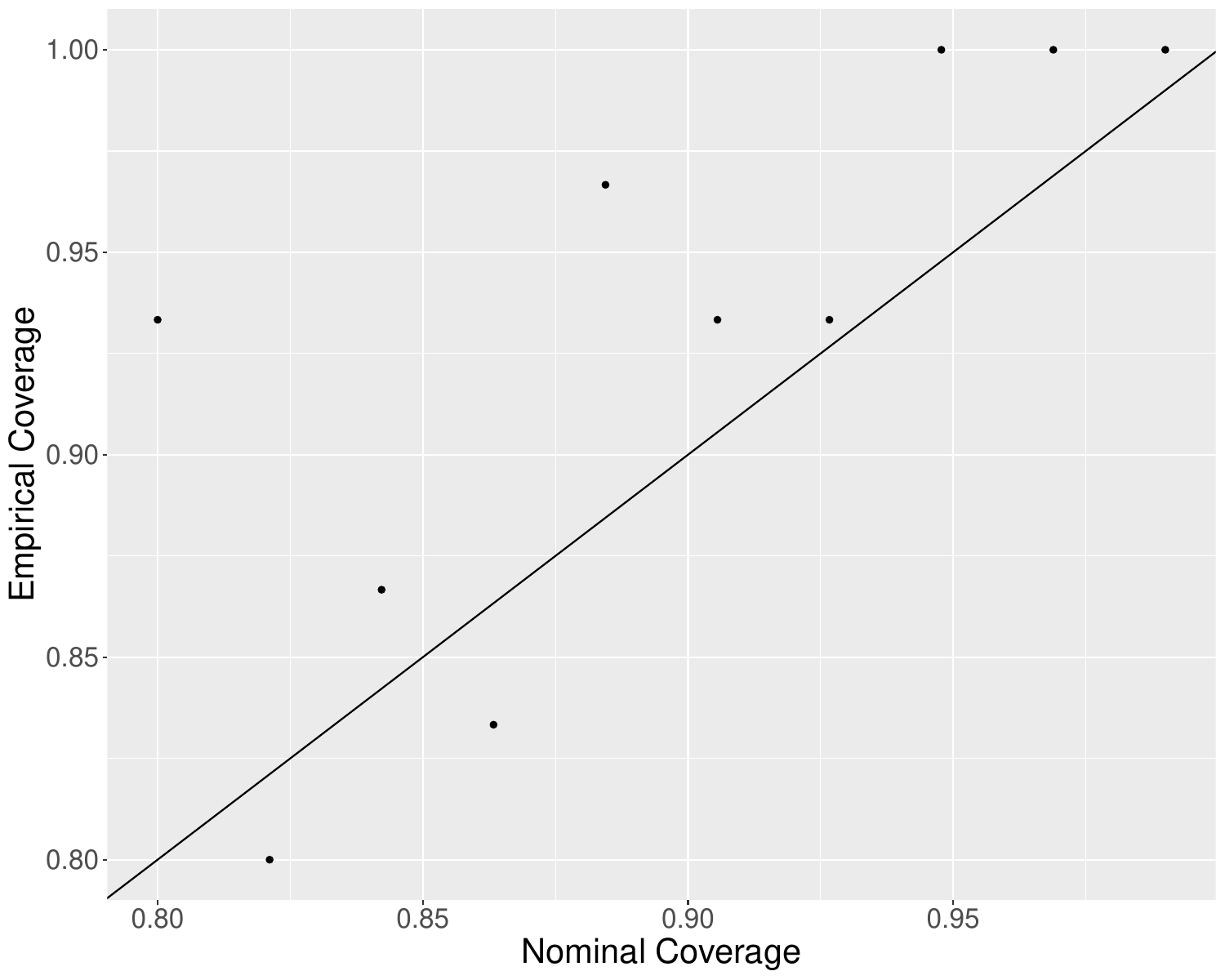}
    \caption{\small{Mass-radius relation: Empirical versus nominal coverage for all regions using only non-conformity scores in the same region as the test point. Note, test points randomly split between nominal coverage levels.}}
    \label{fig:mr_cov}
\end{figure}

Finally, we calculate the coverages in the same way as we did for the simulation setup by calculating the p-value in Algorithm (\ref{alg:cpi}) using $z=\widetilde{M}_i^{\text{obs}}$ for test point $i$ using only the non-conformity scores of points in the same region. We compare our nominal versus empirical coverages in two ways. First, we examine a grid of ten nominal coverages from 80\% to 99\% by randomly choosing test points to be used for each nominal coverage level. Figure \ref{fig:mr_cov} compares these nominal coverages against the empirical coverages when aggregating across all regions. As we can see, our method produces prediction sets with the correct coverages regardless of the nominal coverage set by the practitioner. Second, we examine the empirical coverages across the different regions of the fitted spline for a fixed nominal coverage of 95\% and use all test points. Table \ref{table:mr_cov} shows that we achieve the correct coverage across the regions.

\begin{table}
\centering
\begin{tabular}{c c c} 
 \hline \hline
 Region & Number of Obs. & Empirical Coverage \\ 
 \hline
 1 & 22 & 0.954 \\ 
 2 & 73 & 0.973 \\
 3 & 22 & 1.00 \\
 4 & 88 & 0.989 \\
 5 & 95& 0.989 \\
 \hline \hline
\end{tabular}
\caption{\small{Mass-radius relation: Empirical coverage by region, based on the number of test points in the second column, when producing 95\% prediction intervals. Coverage determined using training observations in the same region as the test point. Note, all test points used.}}
\label{table:mr_cov}
\end{table}

\section{Conclusion} \label{sec:concl}

In this work, we tailor the CP framework discussed in \cite{vovk2005} and \cite{shafer2008tutorial} to the measurement error model. In particular, we demonstrate that the scaled, deleted residual produces approximately exchangeable observations that are robust to a misspecified error distribution, provided the errors are from the class of elliptical distributions. Further, we argue that the distribution of the true covariates does not need to be known or estimated in order to construct these residuals. These properties, thus, allow us to use the CP framework to construct prediction intervals that achieve the specified nominal coverage. Finally, we demonstrate how our method might be used in the astronomy community to construct prediction intervals for exoplanets in which the radius is measured but not the mass. Additionally, while the prediction intervals constructed in this paper only use one covariate value, the methods described can easily be extended to situations with more than one covariate, with or without measurement error.

A major downside of the method used here is that the computation time for constructing these prediction intervals can become infeasible for very large datasets. This is in large part due to calculating the deleted, scaled residuals by individually removing each observation and then refitting the model. In the standard linear regression scenario, one can use one minus the leverage score of the $i$th observation to efficiently calculate the deleted residual of that observation. Further work needs to be done to determine to what degree a similar trick could be used in the measurement error model setting. Alternatively, one could use the inductive conformal method, also known as the split conformal method, which divides the data into training, calibration, and test sets \citep{papadopoulos2008inductive}. However, care is needed to appropriately split the data when the underlying relationship is non-linear, and it is well-known that this method produces larger prediction intervals compared to the transductive CP method described in this paper \citep{vovk2015cross}, due to a reduction in statistical efficiency/power.

\section*{Acknowledgments}

This paper makes use of data from the NASA Exoplanet Archive and made available by the NASA Exoplanet Science Institute at IPAC, which is operated by the California Institute of Technology under contract with the National Aeronautics and Space Administration. 

Further, we would like to acknowledge the computing resources provided by North Carolina State University High Performance Computing Services Core Facility (RRID:SCR\_022168). 

%%%%%%%%%%%%%%%%%%%%%%%%%%%%%%%%%%%%%%%%%%%%%%%%%%
\section*{Data Availability}

The data set was derived from sources in the public domain: The Planetary Systems Table of NASA Exoplanet Archive, at \url{https://doi.org/10.26133/NEA12}. Code and source files to reproduce all figures are publicly available at \url{https://ngierty.github.io/research.html}



\bibliography{refs_prediction} % if your bibtex file is called example.bib

%\begin{thebibliography}{99}
%\end{thebibliography}

%%%%%%%%%%%%%%%%%%%%%%%%%%%%%%%%%%%%%%%%%%%%%%%%%%

%%%%%%%%%%%%%%%%% APPENDICES %%%%%%%%%%%%%%%%%%%%%

\appendix

\section{Notation}

\begin{center}
\begin{tabular}{ c|c }
    \hline \hline
    $\epsilon$ & significance level \\ 
    $x_i$ & true covariate value \\
    $\V{\beta}$ & regression parameter vector \\
    $v_i$ & error from intrinsic scatter with $V(v_i|x_i) = \sigma_{v_i}^2$ \\
    $y_i$ & true response value including $v_i$ \\
    $w_i$ & measurement error on $y$ with $V(w_i|x_i) = \sigma^2_{w_i}$ \\
    $y_i^{obs}$ & observed response value \\
    $q_i$ & measurement error on $x_i$ with $V(q_i|x_i) = \sigma^2_{q_i}$ \\
    $x_i^{obs}$ & observed covariate value \\
    $\tau$ & variance parameter vector for $\sigma_{v_i}^2$ \\
    $\Psi_m$ & set of basis functions transforming $x_i$ \\
    $\eta_i$ & $\sigma_{v_i}^2 + \sigma^2_{w_i} + \V{\beta}_1^\intercal \nabla \Psi_m \nabla \Psi_m^\intercal\V{\beta}_1\sigma_{q_i}^2$ \\
    $\Psi(\M{X})$ & design matrix of the regression relationship \\
    $\M{Q}$ & matrix of errors corresponding to design matrix \\
    $r_i^{obs}$ & non-conformity measure; true unknown residual \\
    $\mu$ & general mean parameter \\
    $\Sigma$ & general variance matrix \\
    $\theta$ & parameter in generalized normal distribution \\
    $\nu$ & degrees of freedom in t-distribution \\
    $\gamma$ & parameter in mass-radius relationship model \\
    $\hat{\cdot}$ & any estimate of a value \\
    \hline \hline
\end{tabular}
\end{center}

\section{Derivation of Moment-Corrected Estimators} \label{sec:beta_hat_deriv}

The derivation of the regression parameter estimator in equation (\ref{eq:beta_hat_mc}) in Section \ref{sec:fglsmem} follows the same derivation as the moment-corrected estimators discussed in \cite{buonaccorsi2010measurement} when using a Taylor series approximation to move from a non-linear measurement error model to an additive measurement error model. We state the details here for completeness and ease of reference for the reader.

We assume that $\{\sigma_{v_i}^2, \sigma_{w_i}^2, \sigma_{q_i}^2\}_{i=1}^n$ are known or can be reasonably estimated. Let $\M{X}$ be the matrix of augmented observations $(1, x_i)$, $\Psi(\M{X})$ the design matrix of the regression relationship in equation (\ref{eq:mem}) without measurement error, $\nabla \Psi(\M{X}) = \diag(\nabla \Psi_m(x_1)^{\intercal}, \dots, \nabla \Psi_m(x_n)^{\intercal})$, $\Psi(\M{X}^{obs})$ the design matrix based on the observed covariates, $\V{y}$ the vector of responses $y_i$, $\V{w}$ the vector of measurement errors on $\V{y}$, $\V{y}^{obs}$ be the vector of observed responses, and $\Sigma_v^{-1} = \diag(\sigma_{v_1}^{-2}, \dots, \sigma_{v_n}^{-2})$. Finally, we let $\M{Q}$ be the matrix of augmented measurement errors on $\Psi(\M{X})$; specifically,
\begin{align*}
    \M{Q} = \begin{bmatrix} 0_m & q_1 \M{I}_m \\ \vdots & \vdots \\ 0_m & q_n \M{I}_m
    \end{bmatrix}
\end{align*} Without measurement error, $\hat{\V{\beta}} = (\Psi(\M{X})^{\intercal} \Sigma_v^{-1} \Psi(\M{X}))^{-1} \Psi(\M{X})^{\intercal} \Sigma_v^{-1} \V{y}$. To correct for the measurement error, we calculate the expected value of the components of $\hat{\V{\beta}}$ and correct for any biases in those values accordingly.

\begin{align*}
    E \big[\Psi(\M{X}^{obs})^{\intercal} \Sigma_v^{-1} \Psi(\M{X}^{obs}) | \M{X} \big] &\approx E \big\{ \big[\Psi(\M{X}) + \nabla \Psi(\M{X}) \M{Q} \big]^{\intercal} \Sigma_v^{-1} \big[\Psi(\M{X}) + \nabla \Psi(\M{X}) \M{Q} \big] | \M{X} \big\} \\
    &= E \big[ \Psi(\M{X})^{\intercal} \Sigma_v^{-1} \Psi(\M{X})  + \Psi(\M{X})^{\intercal} \Sigma_v^{-1} \nabla \Psi(\M{X}) \M{Q} + \M{Q}^{\intercal} \nabla \Psi(\M{X})^{\intercal} \Sigma_v^{-1} \Psi(\M{X}) + \M{Q}^{\intercal} \nabla \Psi(\M{X})^{\intercal} \Sigma_v^{-1} \nabla \Psi(\M{X}) \M{Q} | \M{X} \big] \\
    &= \Psi(\M{X})^{\intercal} \Sigma_v^{-1} \Psi(\M{X}) + \begin{bmatrix}
        0 & 0_m^{\intercal} \\
        0_m & \sum_{i=1}^n \nabla \Psi_m(x_i) \nabla \Psi_m(x_i)^{\intercal} \sigma_{q_i}^2/ \sigma_{v_i}^2
    \end{bmatrix} \\
    &= \Psi(\M{X})^{\intercal} \Sigma_v^{-1} \Psi(\M{X}) + \M{\Sigma}_q
\end{align*}

\begin{align*}
    E \big[\Psi(\M{X}^{obs})^{\intercal} \Sigma_v^{-1} \V{y}^{obs} | \M{X} \big] &\approx  E \big\{ \big[\Psi(\M{X}) + \nabla \Psi(\M{X}) \M{Q} \big]^{\intercal} \Sigma_v^{-1} \big(\V{y} + \V{w} \big) | \M{X} \big\} \\
    &= E \big[ \Psi(\M{X})^{\intercal} \Sigma_v^{-1} \V{y} + \Psi(\M{X})^{\intercal} \Sigma_v^{-1} \V{w} + \\& \M{Q}^{\intercal}  \nabla \Psi(\M{X})^{\intercal} \Sigma_v^{-1} \V{y} + \M{Q}^{\intercal}  \nabla \Psi(\M{X})^{\intercal} \Sigma_v^{-1} \V{w} | \M{X} \big] \\
    &= \Psi(\M{X})^{\intercal} \Sigma_v^{-1} \V{y}
\end{align*}

\noindent
Substituting $\{\sigma_{v_i}^2, \sigma_{w_i}^2, \sigma_{q_i}^2\}_{i=1}^n$ for their initialized values, gives the result in equation (\ref{eq:beta_hat_mc}).

% continuous mapping theorem; slutsky's theorem

\end{document}